\RequirePackage{fix-cm}
\documentclass[smallcondensed]{svjour3}     % onecolumn (ditto)
\smartqed  % flush right qed marks, e.g. at end of proof
\usepackage{graphicx}

\usepackage{cite}
\usepackage[cmex10]{amsmath}
\usepackage{array}
\usepackage{mdwmath}
\usepackage{mdwtab}
\usepackage{eqparbox}
\usepackage{graphics}
\usepackage[varg]{txfonts}
\usepackage{latexsym}
\usepackage{graphicx}
\usepackage{epstopdf}
\usepackage{subfigure}
\usepackage{mathrsfs}
\usepackage{amsmath}
\usepackage{tabularx}

\newcommand{\F}{\mathbb{F}}
\newcommand{\Z}{\mathbb{Z}}

\begin{document}

\title{Hermitian dual-containing constacyclic BCH codes and related quantum codes of length $\frac{q^{2m}-1}{q+1}$%\thanks{Grants or other notes
}


\author{Xubo Zhao$^{1,2,3}$  \and  Xiaoping Li$^{1,2,3}$  \and Qiang Wang$^{4}$
         \and Tongjiang Yan$^{1,2,3}$  %etc.
}


\institute{Xubo Zhao \\
 \email{zhaoxubo@upc.edu.cn}\\
Xiaoping Li\\
\email{xplteaa@163.com}\\
Qiang Wang\\
  \email{wang@math.carleton.ca}\\
Tongjiang Yan\\
\email{yantoji@163.com}\\
1 College of Science, China University of Petroleum (East China), Shandong, Qingdao 266580, China\\
2 Key Laboratory of Applied Mathematics(Putian University), Fujian Province University, Fujian Putian, 351100, China\\
3 Shandong Provincial Key Laboratory of Computer Networks, Shandong, Jinan, 250014, China\\
4 School of Mathematics and Statistics, Carleton University, 1125 Colonel By Drive, Ottawa, Ontario K1S 5B6, Canada
}

\date{Received: date / Accepted: date}

\maketitle
\begin{abstract}
In this paper, we study a family of constacyclic BCH codes over $\mathbb{F}_{q^2}$ of length $n=\frac{q^{2m}-1}{q+1}$, where $q$ is a prime power, and $m\geq2$ an even integer.
The maximum design distance of narrow-sense Hermitian dual-containing constacyclic BCH codes of length $n$ is determined.
Furthermore, the exact dimension of the constacyclic BCH codes with given design distance is computed. As a consequence, we are able to derive the
parameters of quantum codes as a function of their design parameters of the associated constacyclic BCH codes. This improves the result by Yuan et al. (Des Codes Cryptogr 85(1): 179-190, 2017), showing that with the same lengths,
except for three trivial cases ($q=2,3,4$), our resultant quantum codes can always yield strict dimension or minimum distance gains than the ones obtained by Yuan et al..
Moreover, fixing length $n=\frac{q^{2m}-1}{q+1}$, some constructed quantum codes have better parameters or are beneficial complements compared with some known results (Aly et al., IEEE Trans Inf Theory 53(3): 1183-1188, 2007, Li et al., Quantum Inf Process 18(5): 127, 2019, Wang et al., Quantum Inf Process 18(8): 323, 2019, Song et al., Quantum Inf Process 17(10): 1-24, 2018.).

\end{abstract}

\section{Introduction}
\label{intro}
Quantum error correction codes play an important role in protecting quantum information from decoherence in quantum computations and quantum communications. One of the principal problems in quantum error correction is to construct quantum codes with good parameters. After the establishment of the connections between quantum codes and classical codes by
Calderbank et al. \cite{Calderbank1998,ashikhmin2001}, construction of quantum codes can be deduced to find classical ``dual-containing" or ``self-orthogonal"  linear codes with respect to certain inner products.
Among these inner products, the Hermitian inner product has been widely used, and the corresponding Hermitian dual-containing or Hermitian self-orthogonal codes can yield many good quantum codes \cite{Grassl2004,kai2014,chen2015,tang2019, zhang2018,zhang2019, jin2014,wang2015,Hao2018,zhang2015,fang2018,Hu2018,la2014}.

Denote $\mathbb{F}_{q^2}$ be a finite field with $q^2$ elements and $\mathbb{F}_{q^2}^{*}=\mathbb{F}_{q^2}\backslash\{0\}$, where $q$ is a prime power. For any $\alpha \in \mathbb{F}_{q^2}$, the conjugate of $\alpha$ is defined by $\bar{\alpha}=\alpha^{q}$.
Let $\mathbb{F}_{q^2}^{n}$ be the $\mathbb{F}_{q^2}$-vector space of $n$-tuples. A $q^2$-ary linear code $\mathcal{C}$ of length $n$ is an $\mathbb{F}_{q^2}$-subspace of $\mathbb{F}_{q^2}^{n}$. A linear code $\mathcal{C}$ is called an $[n,k,d]_{q^2}$ linear code if its dimension is $k$ and minimum Hamming distance is $d$. Given two vectors $\textbf{u}=(u_{1},u_{2},\cdots,u_{n})$ and $\textbf{v}=(v_{1},v_{2},\cdots,v_{n})\in \mathbb{F}_{q^2}^{n}$, their Hermitian inner product
$\langle \textbf{u, v} \rangle_{H}$ is defined as
\begin{eqnarray}\label{h inner product}
\langle \textbf{u, v} \rangle_{H}=\sum\nolimits_{i=1}^{n}{u_{i}\bar{v_{i}}}.
\end{eqnarray}
The vectors $\textbf{u, v}$ are called orthogonal with respect to the Hermitian inner product if $\langle \textbf{u, v} \rangle_{H}=0$.  For a $q^2$-ary linear code $\mathcal{C}$, its Hermitian dual code is defined as
\begin{eqnarray}\label{h dual code}
\mathcal{C}^{\perp H}=\{\textbf{u}\in \mathbb{F}_{q^2}^{n} \mid \langle \textbf{u, v} \rangle_{H}=0, \forall \textbf{v}\in \mathcal{C}\}.
\end{eqnarray}
 If $\mathcal{C}^{\perp H}\subseteq \mathcal{C}$ and $\mathcal{C}\neq \mathbb{F}_{q^2}^{n}$, $\mathcal{C}$ is called a Hermitian dual-containing code, and $\mathcal{C}^{\perp H}$ is called a Hermitian self-orthogonal code.
One of the most frequently-used and effective construction methods of quantum codes is as follows (see \cite{ashikhmin2001}).
\begin{theorem}\label{h construction}
If $\mathcal{C}$ is an $[n,k,d]_{q^2}$ linear code such that $\mathcal{C}^{\perp H}\subseteq \mathcal{C}$, then there exists an $[[n,2k-n,\geq d]]_{q}$ quantum code.
\end{theorem}
We refer to the construction in Theorem \ref{h construction} as Hermitian construction, which suggests that $q$-ary quantum codes can be obtained from classical Hermitian dual-containing linear codes over $\mathbb{F}_{q^2}$.

Bose-Chaudhuri-Hocquenghem (BCH) codes are a significant class of good cyclic codes, which have efficient encoding and decoding algorithm and strong error-correcting capability. They have been widely employed in data storage systems and satellite communications. Another important application of BCH codes is to construct quantum codes. Aly et.al \cite{Aly2006, Aly2007} studied Euclidean or Hermitian dual-containing BCH codes of general length, they derived a formula for the dimension of narrow-sense BCH codes with small design distance, and then specified the parameters of quantum BCH codes in terms of the design distance. Thereafter, a lot of good quantum codes have been obtained by BCH codes \cite{la2009,la2014,Hao2018,liuyang2019,tang2019,zhang2018,zhang2019}.

Constacyclic BCH codes \cite{la2016} as a generalization version of the well-known BCH codes, are naturally considered to construct quantum codes.
Generally speaking, the parameter of quantum codes constructed from constacyclic BCH codes \cite{liu2017,wang2020,zhu2018,lin2004,Yuan2017,kai2018,wang2019,guo2020} might be better than the ones derived from BCH codes.
Based on Hermitian construction, in order to construct quantum codes, it is necessary to know the exactly dimension and minimum distance of the classical codes. However, for general code length, the dimension and minimum distance of BCH or constacyclic BCH codes have limited knowledge
because of their intricate structure \cite{charpin1998,la2014}. Therefore, it is very hard to obtain the precise dimension and minimum distance of quantum codes in general, a summary of some known constructions of quantum constacyclic codes with special code lengths is provided in Table 3 of literature \cite{wang2020}.

In \cite{zhu2018}, Zhu et al. studied the dimension, the minimum distance, and the weight distribution of certain negacyclic BCH codes (a subclass of constacyclic BCH codes) of length $n=\frac{q^{2m}-1}{q-1}$ over $\mathbb{F}_{q}$ with odd $q$ and any positive integer $m$. From this class of negacyclic BCH codes, they acquired some quantum codes with good parameters. In \cite{kai2018} and \cite{guo2020}, Kai and Guo et al. investigated negacyclic BCH codes of length $n=\frac{q^{2m}-1}{2}$ over $\mathbb{F}_{q}$ with odd $q$, or odd $q$ and odd $m$, respectively. Later, Wang et al. \cite{wang2020} discussed narrow-sense and non-narrow-sense negacyclic BCH codes of length $n=\frac{q^{2m}-1}{a}$, where $a|(q^m-1)$ is even, both $q$ and $m$ are odd.
Then they constructed some quantum codes with good parameters from those dual-containing negacyclic BCH codes.
By investigation of suitable properties on $q^2$-ary cyclotomic cosets, Liu et al. \cite{liu2017} determined a class of Hermitian dual-containing constacyclic BCH codes of length $n=q^{2m}+1$ over $\mathbb{F}_{q^2}$ with odd $q$ and any integer $m\geq2$. Then they obtained some new quantum codes with minimum distance $d>2q^2$. Wang et al. \cite{wang2019} focused on a class of $q^2$-ary constacyclic BCH codes of length $n=2(q^{m}+1)$ with
odd $q$ and any $m\geq3$, via Hermitian construction, they gained many new quantum codes, which are not covered in the literature.

As shown above, for the case of odd $q$, many $q^2$-ary constacyclic BCH codes with special code lengths have been well studied
and have been used to construct quantum codes with desirable parameters via Hermitian construction. While for the case of even $q$, especially with even $m$, Hermitian dual-containing constacyclic BCH codes are much less investigated due to the difficulty of computing the exact dimension of this class of codes.
Based on classical quaternary constacyclic BCH codes, Lin \cite{lin2004} got a class of binary quantum codes of length $\frac{4m-1}{3}$ ($m\geq4$ is an integer) with nice parameters.
Later, Yuan et al. \cite{Yuan2017} generalized the binary quantum constacyclic codes construction of \cite{lin2004} to the $q$-ary case with length $n=\frac{q^{2m}-1}{q+1}$ by applying classical $q^2$-ary constacyclic BCH codes, where $q$ is a prime power and $m\geq2$. Moreover, compared with the corresponding ones in \cite{lin2004}, the narrow-sense constacyclic BCH codes in \cite{Yuan2017} have relatively large design distance. To determine the dimension of the Hermitian dual-containing constacyclic BCH codes, Yuan et al. followed a directly generalized formula (see Eq.(3) in \cite{Yuan2017}) of \cite{lin2004}.
However, the dimension of narrow-sense constacyclic BCH codes presented in \cite{Yuan2017} are just the lower bound of their actual
dimension. The main reason is that the union of $q^2$-cyclotomic cosets for the defining set of constacyclic BCH codes will involve more repeated ingredients as the size of finite field and the design distance become larger. Very recently, Wang et al. \cite{wang2019} study Hermitian dual-containing narrow-sense constacyclic BCH codes of length $\frac{q^{2m}-1}{\rho}$ and aim at constructing new quantum codes
with the corresponding constacyclic BCH codes, where $\rho|(q+1)$, and $m\geq 3$. By employing the technique of \cite{ding2017}, they determine all coset leaders of cyclotomic cosets in the defining set and discuss the $q$-adic expansion about the design distance to compute the dimension of constacyclic BCH codes. This method is efficient for the case of odd $m$ (see Theorem 3 in \cite{wang2019}), however, the derived parameters of constacyclic BCH codes and the related quantum codes are not concise enough for the case of even $m\geq4$ (see Theorem 4 in \cite{wang2019}).

We found that some optimal or almost optimal codes can be obtained from the narrow-sense constacyclic BCH codes of length $n=\frac{q^{2m}-1}{q+1}$
with special values of $q$ and $m$ (see Remark 2 in Subsection \ref{Quantum Codes Construction}). In addition, some quantum codes constructed from narrow-sense constacyclic BCH codes of length $n=\frac{q^{2m}-1}{q+1}$ have better parameters compared to some known results studied in the literature \cite{Yuan2017,Aly2007,ruihu2019,edel2019,wang2019,Hao2018}. Thus, this family of constacyclic BCH codes deserve to be investigated further.
In this article, we focus on determining the parameters of Hermitian dual-containing constacyclic BCH codes and the related quantum codes of length
$n=\frac{q^{2m}-1}{q+1}$ for the case of even $m$, and the results about the case of odd $m$ can be found in Subsection 2.2 of \cite{Yuan2017} or Theorem 3 in \cite{wang2019}. Note that in the case of even $m$, it is more challenging to derive the parameters of this family of Hermitian dual-containing constacyclic BCH codes, since the defining set of the considered codes involves a more intricate structure compared with the case of odd $m$.
We provide necessary and sufficient conditions for the existence of narrow-sense Hermitian dual-containing constacyclic BCH codes (see Theorem \ref{exact Hermitian dual}), which allows one to identify all constacyclic BCH codes that can be used in quantum codes construction.
Furthermore, different from the idea of finding out all coset leaders of the union of cyclotomic cosets in the defining set, we counting the joint
cyclotomic cosets via solving the associated congruent equations. Moreover,
based on a detailed study of properties of the $q^2$-cyclotomic cosets (see Lemmas \ref{s cardinality},
\ref{exact cardinalities}), we compute the exact dimensions of this family constacyclic BCH codes for all design distance $\delta $ in the range $2\leq \delta \leq  \delta^{max}$ (see Theorem \ref{classic Hermitian dual containing linear codes}).
These results facilitate us to determine the parameters of quantum codes as a function of their design parameters $n$, $q$, and $\delta$ of the related narrow-sense constacyclic BCH codes (see Theorem \ref{quantum codes}).

The remainder of this paper is organized as follows. In Section \ref{Preliminaries},
some basic definitions and properties about constacyclic codes are reviewed.
In Section \ref{Hermitian Dual-containing Conditions And The Dimension}, the Hermitian dual-containing conditions, as well as properties of corresponding cyclotomic cosets of narrow-sense constacyclic BCH codes are investigated. Some new quantum codes
are constructed from the underlying constacyclic BCH codes, and the parameters of resultant quantum codes are compared with previous results. The conclusion of this paper is given in  Section \ref{Summary}.

\section{Preliminaries}\label{Preliminaries}
In this section, we review some basic notation and results about quantum codes and constacyclic codes. Throughout this paper, $\mathbb{F}_{q^2}$ denotes the finite field with $q^2$ elements, where $q$ is a prime power, $\mathbb{Z}$ represents the ring of integers, $\lfloor X \rfloor$ means the maximal integer, which is not more than $X$, and $\lceil X \rceil$ is the minimal integer, which is not less than $X$. The bracket notation [statement] takes the value 1 if statement is true, and 0 otherwise, for instance, we have [$q$ even]=$q-1$ (mod 2).
%\subsection{Constacyclic codes}\label{sub Constacyclic Codes}
%Let $\mathbb{F}_{q^2}^{*}$ be the multiplicative group of nonzero elements of $\mathbb{F}_{q^2}$.

For $\eta \in \mathbb{F}_{q^2}^{*}$ , a linear code $\mathcal{C}$ of length $n$ over $\mathbb{F}_{q^2}$ is called $\eta$-\emph{constacyclic code}\cite{Berlekamp1968,minjia2016} provided that for each codeword $\textbf{c}=(c_{0},c_{1},\cdots,c_{n-1})$ in $\mathcal{C}$, $(\eta c_{n-1},c_{0},\cdots,c_{n-2})$ is also a codeword in $\mathcal{C}$. Some of the most important classes of codes can be realized as special cases of constacyclic codes. For example, the case $\eta=1$ gives \emph{cyclic codes}, and $\eta=-1$ yields \emph{negacyclic codes}.
Customarily, each codeword $\textbf{c}=(c_{0},c_{1},\cdots,c_{n-1})\in \mathcal{C}$ is identified with its polynomial representation
$c(x)=c_{0}+c_{1}x+\cdots+c_{n-1}x^{n-1}$. It is easy to check that an $\eta$-constacyclic code $\mathcal{C}$ of length $n$ over $\mathbb{F}_{q^2}$ corresponds to a principal ideal $\langle g(x) \rangle$ of the quotient ring $\mathbb{F}_{q^2}[x]/\langle x^n-\eta\rangle$, where $g(x)$ is a monic divisor of $x^n-\eta$. In this case, $\mathcal{C}$ is generated uniquely by $g(x)$, i.e., $\mathcal{C}=\langle g(x)\rangle$, the polynomial $g(x)$ is called the \emph{generator polynomial} of the code $\mathcal{C}$, and the dimension of $\mathcal{C}$ is $n-$deg$(g(x))$.

 We assume that $n$ and $q$ are relatively prime, i.e., gcd$(n,q)=1$, so that the polynomial $x^n-\eta$ over $\mathbb{F}_{q^2}$ does not involve repeated roots. We denote the order of $\eta$ in the multiplicative group $\mathbb{F}_{q^2}^{*}$ by ord($\eta$).
 Assume that ord($\eta)=r$, then $\eta$ is called a \emph{primitive} $r$th \emph{root of unity}.
 Suppose that the multiplicative order of $q^2$ modulo $nr$ is $m$ (the smallest positive $m$ such that $nr$ divides $q^{2m}-1$), namely, ord$_{nr}(q^2)=m$. Then there exists a primitive $nr$th root $\beta$ of unity
 in $\mathbb{F}_{q^{2m}}$ such that $\beta^{n}=\eta$. Let $\xi=\beta^{r}$, then $\xi$ is a primitive $n$th root of unity. Thus, the roots of $x^{n}-\eta$ are $\beta\xi^{i}=\beta^{1+r i}$, $i=0, 1,\cdots, n-1$, i.e., $x^{n}-\eta=\prod_{i=0}^{n-1}(x-\beta^{1+r i})$.
 Let
 \begin{eqnarray}\label{O}
\mathcal{O}=\{1+r i \mid 0\leq i \leq n-1\}\ \  (\textrm{mod}\ nr).
\end{eqnarray}
The \textit{defining set }of the $\eta$-constacyclic code $\mathcal{C}=\langle g(x) \rangle$ is defined as
\begin{eqnarray}\label{T}
T=\{j\in\mathcal{O} \mid g(\beta^{j})=0\}.
\end{eqnarray}
Let $C_{s}$ be the $q^2$-cyclotomic cosets modulo $nr$, which
contains $s$. That is $C_{s}=\{sq^{2l}\ \textrm{mod}\ nr\mid l\in Z, 0\leq l \leq m_{s}-1\}$, where $m_{s}$ is the smallest positive integer such that $sq^{2m_{s}}\equiv s\ \textrm{mod}\ nr$, and $s$ is not necessarily the smallest number in the coset $C_{s}$.
Recall that for each integer $\tau$, with $0\leq \tau < nr-1$, the minimal polynomial of $\beta^{\tau}$ over ${\F}_{q^2}$
is $M_{{\tau}}(x)=\prod_{i\in C_{\tau}}(x-\beta^{i})$. And $x^{nr}-1=\prod_{\tau\in \Omega}M_{{\tau}}(x)$ is the factorization of $x^{nr}-1$ into
irreducible factors over ${\F}_{q^2}$, where $\Omega$ is the set of representatives of all the distinct $q^2$-cyclotomic cosets modulo $nr$ (see \cite{Aydin2001,Huffman2003}). Obviously, $(x^{n}-\eta)|(x^{nr}-1)$, then $x^{n}-\eta=\prod_{\tau\in \mathcal{O}\cap\Omega}M_{{\tau}}(x)$, thereby the defining set of code $\mathcal{C}=\langle g(x) \rangle$ must be a union of some $q^2$-cyclotomic cosets modulo $nr$.

 \begin{definition}\label{Constacyclic BCH code}\cite{la2016,ruihu2019}
 Assume that gcd$(n,q)=1$. Let $\mathcal{C}=\langle g(x) \rangle$ be an $\eta$-constacyclic code of length $n$ over $\mathbb{F}_{q^2}$, where $\eta$ is a primitive $r$th root of unity. Let $\beta$ be a primitive $nr$th root of unity in some extension field of $\mathbb{F}_{q^2}$, such that $\beta^{n}=\eta$. An $\eta$-constacyclic BCH code of length $n$ with design distance $\delta$ is an $\eta$-constacyclic code with defining set
\begin{eqnarray}\label{CBCH generator}
 T=\cup_{i=0}^{\delta-2}C_{b+r i},
\end{eqnarray}
where $b=1+jr, j\in \Z$. Such a code is called primitive if $n=q^{2m}-1$, and non-primitive, otherwise. If $b=1$, $\mathcal{C}$ is called a narrow-sense code, and non-narrow-sense, otherwise.
 \end{definition}

 For an $\eta$-constacyclic BCH code $\mathcal{C}$, the following BCH bound (see \cite{Aydin2001} Theorem 2.2 or \cite{chen2015} Theorem 2.8) shows the relationship between the minimum distance of $\mathcal{C}$ and the design distance $\delta$ of $\mathcal{C}$.

\begin{theorem}\label{BCH bound}
 Let $\mathcal{C}$ be an $\eta$-constacyclic BCH code with defining set $T\subseteq \mathcal{O}$.
 If $\{1+r i \mid a\leq i \leq a+\delta-2\}\subseteq T$, where  $a\in \Z$. Then the minimum distance of $\mathcal{C}$ is at least $\delta$.
\end{theorem}

%\subsection{Hermitian construction}
 As mentioned in Theorem \ref{h construction}, $q$-ary quantum codes can be obtained from classical Hermitian dual-containing linear codes over $\F_{q^{2}}$ by Hermitian construction. Normally, the condition of  Hermitian dual-containing  $\mathcal{C}^{\perp H}\subseteq \mathcal{C}$ can be guaranteed by the popular defining set criterion in the following (for example, see \cite{kai2014} Lemma 2.2).
\begin{lemma}\label{dual containing iff condition}
Let $\mathcal{C}=\langle g(x) \rangle$ be an $\eta$-constacyclic code of length $n$ over $\mathbb{F}_{q^2}$ with defining set
$T$. Then $\mathcal{C}$ contains its Hermitian dual code, namely, $\mathcal{C}^{\perp H}\subseteq \mathcal{C}$,
if and only if $T\cap T^{-q}=\emptyset$, where $T^{-q}=\{-qj \ \textrm{mod}\  nr \mid j\in T\}$.
\end{lemma}

Chen et.al in \cite{chen2015} have shown that if an $\eta$-constacyclic BCH code over $\mathbb{F}_{q^2}$ is a Hermitian dual-containing code, then we have ord$(\eta)|(q+1)$. From now on till the end of this paper, we let $n=\frac{q^{2m}-1}{q+1}$, where $m\geq2$ is an even, $q$ is a prime power. We take $\eta=\alpha^{q-1}$ as in \cite{Yuan2017}, where $\alpha$ is a primitive element of $\mathbb{F}_{q^2}$.  Thus, $r=$ord$(\eta)=q+1$, $nr=q^{2m}-1$, and $T^{-q}=\{-qj \ \textrm{mod}\  q^{2m}-1\mid j\in T\}$.
\section{Construction of quantum codes from constacyclic BCH codes}\label{Hermitian Dual-containing Conditions And The Dimension}
In this section, we first determine the corresponding parameters for which $\mathcal{C}^{\perp H}\subseteq \mathcal{C}$, and then derive the dimensions and bound the minimum distances of $\mathcal{C}$, thereby we can construct related quantum codes from constacyclic BCH codes $\mathcal{C}$.

\subsection{Hermitian dual-containing conditions and the dimension of narrow-sense constacyclic BCH codes of length $n=\frac{q^{2m}-1}{q+1}$}\label{narrow Equivalent Conditions Of Hermitian Dual-containing}
%In this subsection, we are interested in obtaining the criterias of the existence of Hermitian dual-containing for $\eta$-constacyclic codes.
Yuan et.al in \cite{Yuan2017} gave a sufficient condition on the design distances for which a narrow-sense constacyclic BCH code is Hermitian
dual-containing code.
In the following Theorem, we can derive a necessary and sufficient condition on the design distance for which a narrow-sense constacyclic BCH code is Hermitian dual-containing code. The significance of this result is that it allows us to identify all constacyclic BCH codes that can be used for construct quantum codes.
\begin{theorem}\label{exact Hermitian dual}
Let $n=\frac{q^{2m}-1}{q+1}$, where $q$ is a prime power, and $m\geq 2$ is an even. Denote $\eta\in \mathbb{F}^*_{q^2}$, and ord$(\eta)=r=q+1$. Let $\mathcal{C}=\langle g(x) \rangle$ be a narrow-sense constacyclic BCH code with defining set $T=\bigcup_{i=0}^{\delta-2}C_{1+ri}$, where $C_{1+ri}$, $0\leq i\leq \delta-2$, are $q^2$-cyclotomic cosets  modulo $nr$, and $\delta\in \mathbb{Z}$.
Then $\mathcal{C}^{\perp H}\subseteq \mathcal{C}$,
if and only if $\delta$ is in the range of
\begin{eqnarray}\label{the range of tau}
2\leq \delta \leq \delta^{max},
\end{eqnarray}
where
\begin{eqnarray}\label{max delta}
\delta^{max}=\left\{\begin{array}{ll}
\frac{q^{3}-q^2+q+3}{q+1}+1, & \ \ m=2\ \textrm{and}\ q\geq 5,\\
\frac{q^{m+1}-q^2+q+3}{q+1}, & \ \ \textrm{otherwise}.
\end{array}\right.
\end{eqnarray}
\end{theorem}

\emph{Proof}
Firstly, we prove the sufficiency.
In terms of Lemma\ref{dual containing iff condition}, $\mathcal{C}^{\perp H}\subseteq \mathcal{C}$ if and only if $T\cap T^{-q}=\emptyset$.
Assume that $T\cap T^{-q} \neq \emptyset$, then there exist two integers $i, j$, $0\leq i, j \leq \delta^{max}-2$, such that
\begin{eqnarray}\label{X1X2}
1+rj \equiv -q(1+ri)q^{2l} \ (\textrm{mod}\ nr),
\end{eqnarray} or equivalently,
\begin{eqnarray}\label{X1X22}
(1+rj)q^{2(m-l)} \equiv -q(1+ri) \ (\textrm{mod}\ nr),
\end{eqnarray} where $l\in \{0,1,\cdots,m-1\}$. We note that $m\geq 2$ is even, and if $\frac{m}{2}\leq l \leq m-1$,
then $0\leq m-l-1 \leq \frac{m}{2}-1$.
Thus we just need to consider the following equation,
\begin{eqnarray}\label{X1X2equation11}
(1+ri)q^{2l+1}+1+rj \equiv 0 \ (\textrm{mod}\ nr),\ l\in \{0,1,\cdots,\frac{m}{2}-1\}.
\end{eqnarray}

 If $m>2, \textrm{or}\  m=2\ \textrm{and}\ q\leq 4$, then $1\leq 1+rj\leq q^{m+1}-q^2-q+2<q^{m+1}-1$, $q\leq(1+ri)q^{2l+1}\leq q^{m-1}(q^{m+1}-q^2-q+2)\leq q^{2m}-q^{m+1}$ $(0\leq l \leq \frac{m}{2}-1)$. If $m=2 \ \textrm{and}\ q\geq 5$, then $1\leq 1+rj\leq q^{3}-q^2+3$, $q\leq(1+ri)q^{2l+1}\leq q(q^{3}-q^2+3)=q^{4}-q^{3}+3q$ $(l=0)$. For both cases, we can get $1+q\leq(1+ri)q^{2l+1}+1+rj<q^{2m}-1=nr$, hence $(1+ri)q^{2l+1}+1+rj \not\equiv 0 \ (\textrm{mod}\ nr)$. This
contradicts to the assumption $T\cap T^{-q} \neq \emptyset$.
So we conclude that if $0\leq i, j \leq \delta^{max}-2$, then $T\cap T^{-q}=\emptyset$, namely, $\mathcal{C}^{\perp H}\subseteq \mathcal{C}$.

Now we will show the necessity, namely, if $\delta$ exceeds $\delta^{max}$, then $\mathcal{C}^{\perp H}\not\subseteq \mathcal{C}$.
If $m>2, \textrm{or}\  m=2\ \textrm{and}\ q\leq 4$, assume that $\delta_{0}=\delta^{max}+1$, i.e., $\delta_{0}-2=\frac{q^{m+1}-q^2+2}{q+1}$, then we have
\begin{eqnarray}\label{contradiction 1}
-q[1+(q+1)\frac{q^{m+1}-q^2+2}{q+1}]q^{m-2} &\equiv& q^{m+1}-3q^{m-1}-1 \nonumber\\
&\equiv& 1\!+\!(q+1)i_{0}\ (\textrm{mod}\ nr),
\end{eqnarray}
where $i_{0}=\frac{q^{m+1}-3q^{m-1}-2}{q+1}$.
If $m=2 \ \textrm{and}\ q\geq 5$, assume that $\delta_{0}=\delta^{max}+1$, i.e., $\delta_{0}-2=\frac{q^{3}-q^2+q+3}{q+1}$, then we have
\begin{eqnarray}\label{contradiction 2}
-q[1+(q+1)\frac{q^{3}-q^2+q+3}{q+1}] &\equiv& q^{3}-q^{2}-4q-1 \nonumber\\
&\equiv& 1\!+\!(q+1)i_{0}\ (\textrm{mod}\ nr),
\end{eqnarray}
where $i_{0}=\frac{q^{3}-q^{2}-4q-2}{q+1}$. Note that for both cases, $0\leq i_{0}\leq\delta_{0}-2$, so $1+ri_{0}\in T\cap T^{-q}$. Thus, $T\cap T^{-q}$ is not empty, so $\mathcal{C}^{\perp H}\not\subseteq \mathcal{C}$.
The desired result follows.\qed

%\subsection{The Dimension Of Narrow-Sense $\eta$-Constacyclic BCH Codes Of Length $n=\frac{q^{2m}-1}{q+1}$}\label{The Dimension Of  %Constacyclic BCH Codes}
Some counting properties about
$q^2$-cyclotomic cosets $C_{1+ri}$ are provided in the following lemma, which is essential to our subsequent arguments.
% which will be instrumental for our main results
\begin{lemma}\label{s cardinality}
We keep the notation of Theorem \ref{exact Hermitian dual}. The cardinalities of $q^2$-cyclotomic cosets $C_{1+ri}$ modulo $nr$, with $i$ in the range $0\leq i \leq \frac{q^{m+1}-2}{q+1}$, can be computed as follows,
\begin{eqnarray}\label{cyclotomic cosets cardinality}
| C_{1+ri} |=\left\{\begin{array}{ll}
m/2, & \ \ q>2\ \textrm{is}\ \textrm{even},\ \textrm{and}\ i=i^{*},\\
m, & \ \ otherwise,\\
\end{array}\right.
\end{eqnarray}
where $i^{*}=\frac{\lceil \frac{q+1}{2}\rceil(q^{m}+1)-[q\ even]}{q+1}$.
\end{lemma}
\emph{Proof}
Note that ord$_{nr}(q^2)=m$, so we have $|C_{1+ri}|\ \big|\  m$, $0\leq i\leq \frac{q^{m+1}-2}{q+1}$.
Suppose that for some $i$, $| C_{1+ri} |<m$, where $0\leq i \leq \frac{q^{m+1}-2}{q+1}$. Then there exists a positive integer $l$, $1\leq l< m$,
such that $(1+ri)q^{2l}\equiv 1+ri\ (\textrm{mod}\  nr)$, namely,
\begin{eqnarray}\label{seek contradiction}
(1+ri)(q^{2l}-1)\equiv 0\ (\textrm{mod}\ nr ).
\end{eqnarray}
Since $m\geq2$ is an even, then $| C_{1+ri} |\leq\frac{m}{2}$, and $1\leq l\leq \frac{m}{2}$ in Eq.(\ref{seek contradiction}).
We first consider the case of $1\leq l\leq \frac{m}{2}-1$ $(m\geq4)$. In this case, since $1\leq1+ri\leq 1+(q+1)\frac{q^{m+1}-2}{q+1}=q^{m+1}-2+1<q^{m+1}$,
$q^2-1\leq q^{2l}-1\leq q^{m-2}-1$, we have $q^2-1\leq(1+ri)(q^{2l}-1)<q^{m+1}(q^{m-2}-1)<q^{2m-1}<nr=q^{2m}-1$. Hence
$(1+ri)(q^{2l}-1)\not\equiv 0\ (\textrm{mod}\ nr )$, contradicting the assumption $| C_{1+ri} |<m$.
And then we consider the case of $l=\frac{m}{2}$ $(m\geq2)$. In this case, Eq.(\ref{seek contradiction})
becomes $(1+ri)(q^{m}-1)\equiv 0\ (\textrm{mod}\ nr )$, which is equivalent to
\begin{eqnarray}\label{s last q+1}
(q+1)i\equiv -1\ (\textrm{mod}\ q^m+1).
\end{eqnarray}
Note that gcd($q+1,q^m+1$)=gcd($q+1,q-1$), thus, gcd($q+1,q^m+1$)=1 if $q$ is even, and gcd($q+1,q^m+1$)=2 if $q$ is odd. In terms of \cite{kumanduri1998} Proposition 3.2.7, when $q$ is odd, Eq.(\ref{s last q+1}) has no solution, when $q$ is even, Eq.(\ref{s last q+1}) has exactly one solution, and
the unique solution can be given by
\begin{eqnarray}\label{s i expression}
i=i^{*}\triangleq\frac{\lceil \frac{q+1}{2}\rceil(q^{m}+1)-[q\ even]}{q+1}.
\end{eqnarray} It is easy to see that $i^{*}<\frac{q^{m+1}-2}{q+1}$ for even $q$, $q>2$, and $i^{*}>\frac{q^{m+1}-2}{q+1}$ for $q=2$.
Thus, the proof is complete.\qed
\textbf{Remark 1.} When $i$ in the range $0\leq i \leq \delta^{max}-2$, then for even $q$,
if $q=2$, or $q=4$ and $m=2$, we have $i^{*}>\delta^{max}-2$.
Therefore, if $0\leq i \leq \delta^{max}-2$, then $| C_{1+ri^{*}} |=m/2$, only if $q=4, m>2$, or
$q>4$ is even.

The following lemma derives the cardinality of $\bigcup_{i=0}^{\delta-2}C_{1+ri}$, $2\leq\delta\leq \delta^{max}$, which contributes to computing the precise dimensions of narrow-sense constacyclic BCH codes with defining set $T=\bigcup_{i=0}^{\delta-2}C_{1+ri}$. To compute
$| \bigcup_{i=0}^{\delta-2}C_{1+ri} |$, we
entail the following tasks:\\
(1). count every cardinality of $q^2$-cyclotomic coset $C_{1+ri}$, $0\leq i \leq \delta-2$, which has done in Lemma \ref{s cardinality},\\
(2). find out all disjoint or joint $q^2$-cyclotomic cosets in $\bigcup_{i=0}^{\delta-2}C_{1+ri}$.\\
Generally, compared with the case of odd $m$, task (2) is more challenging in the case of even $m$, especially $q$ is also even, since
the disjoint or joint $q^2$-cyclotomic cosets are difficult to determine, and the sizes of $q^2$-cyclotomic cosets are not always the same.
Different from the idea of determining all coset leaders of the union of cyclotomic cosets in the defining set, we counting the joint
cyclotomic cosets via solving the associated congruent equations. Consequently, $| \bigcup_{i=0}^{\delta-2}C_{1+ri} |$ is a function of
$n$, $q$, and $\delta$.
%As mentioned in \cite{Aly2007}, for a BCH code, if the multiplicative order $m$ of $q$ modulo $n$ is larger than 1, then the defining set of the code has a more intricate structure, so proofs become more involved.
\begin{lemma}\label{exact cardinalities}
We keep the notation of Theorem \ref{exact Hermitian dual}, then
\begin{eqnarray}\label{even m case 1}
| \bigcup_{i=0}^{\delta-2}C_{1+ri} |&\!\triangleq\!&\mathscr{N}(q,m,\delta)\\
&\!=\!&\left\{\begin{array}{ll}
(\delta-N_{1}-1)m,&\ 2\leq \delta<i^{*}+2,\\
(\delta-N_{1}-1-\frac{1}{2}[q\geq4\ even])m,&\ i^{*}+2\leq \delta<i^{*}+2+\frac{q^m-1}{q+1},\\
(\delta-N_{1}-1-\frac{1}{2}[q\geq4\ even]-N_{2})m,&\ i^{*}+2+\frac{q^m-1}{q+1}\leq \delta\leq \delta^{max},
\end{array}\right.
\end{eqnarray}
where $N_{1}=\lfloor \frac{\delta-(q+1)}{q^2}\rfloor+1$, and
\begin{eqnarray}\label{oeeqequihold1}N_{2}=\left\{\begin{array}{ll}
\lfloor\frac{(q+1)(\delta-3)}{q^{m}-1}\!-\!\lceil\frac{q+1}{2}\rceil\rfloor,&\ i^{*}\!+\!2+\frac{q^m-1}{q+1}\leq\delta\leq\delta^{max},\  q\geq 5, \\
0, &\ otherwise.
\end{array}\right.
\end{eqnarray}
\end{lemma}
\emph{Proof}
Assume that $C_{1+ri}=C_{1+r\bar{i}}$, where $i, \bar{i}\in \Z$,
and $0\leq i<\bar{i}\leq \delta^{max}-2$. Then there exists an integer $l$, $1\leq l \leq m-1$, such that
\begin{eqnarray}\label{m even equ1}
1+r\bar{i}\equiv(1+ri)q^{2l}\ (\textrm{mod}\ q^{2m}-1).
\end{eqnarray}
Since $\textrm{gcd}(q^{m}, q^{2m}-1 )=1$, Eq.(\ref{m even equ1}) is equivalent to
\begin{eqnarray}\label{m even equ2}
(1+r\bar{i})q^{2(m-l)}\equiv1+ri\ (\textrm{mod}\ q^{2m}-1).
\end{eqnarray}
Thus, if $m\geq4$ , Eq.(\ref{m even equ1}) can be equivalently transformed to the following two equations,
\begin{eqnarray}\label{m even equ3}
1+r\bar{i}\equiv(1+ri)q^{2l}\ (\textrm{mod}\ q^{2m}-1), 1\leq l \leq m/2,
\end{eqnarray}
and
\begin{eqnarray}\label{m even equ4}
(1+r\bar{i})q^{2l'}\equiv 1+ri\ (\textrm{mod}\ q^{2m}-1), 1\leq l' \leq m/2-1.
\end{eqnarray}

Note that for $m=2$, Eq.(\ref{m even equ1}) can be reduced to a special case of Eq.(\ref{m even equ3}), namely, the value of $l$ takes $m/2$ in
Eq.(\ref{m even equ3}). Also note that if $m\geq4$, for Eq.(\ref{m even equ4}), we have $1+(q+1)\leq 1+r\bar{i} \leq 1+r(\delta^{max}-2)<q^{m+1}-1$,
and $q^2(q+2)\leq (1+r\bar{i})q^{2l'}<q^{2m}-1$. Thus, by Eq.(\ref{m even equ4}), one can deduce $(1+r\bar{i})q^{2l'}=1+ri$ for $1\leq l' \leq m/2-1$, but this contradicts to the assumption $i<\bar{i}$. Therefore,
for Eq.(\ref{m even equ1}), we just need to consider the formulas in Eq.(\ref{m even equ3}).

For Eq.(\ref{m even equ3}), if $1\leq l \leq m/2-1$ ($m\geq 4$), we have $1\leq 1+ri \leq 1+r(\delta^{max}-3)<q^{m+1}-q^2$,
and $q^2\leq (1+ri)q^{2l}<(q^{m+1}-q^2)q^{m-2}=q^{2m-1}-q^m<q^{2m}-1$. Thus, in the case of $1\leq l \leq m/2-1$ ($m\geq 4$), the eligible solutions of Eq.(\ref{m even equ3}) should satisfy
\begin{eqnarray}\label{m even equ5}
1+r\bar{i}= (1+ri)q^{2l},
\end{eqnarray}
namely,
\begin{eqnarray}\label{m even equ5}
\bar{i}= \frac{q^{2l}-1}{q+1}+iq^{2l}=(q-1)[q^{2(l-1)}+q^{2(l-2)}+\cdots+q^2+1]+iq^{2l},
\end{eqnarray}
which implies
\begin{eqnarray}\label{m even equ6}
\bar{i}\equiv q-1\ (\textrm{mod}\ q^2).
\end{eqnarray}
For the case of $l=m/2$, Eq.(\ref{m even equ3}) becomes
\begin{eqnarray}\label{m even equ7}
1+r\bar{i}\equiv(1+ri)q^{m}\ (\textrm{mod}\ q^{2m}-1),
\end{eqnarray}
or equivalently,
\begin{eqnarray}\label{m even equ8}
r\bar{i} &\equiv& ri+(q^m-1)(1+ri)\ (\textrm{mod}\ q^{2m}-1),\\
\Longleftrightarrow \bar{i} &\equiv& i+\frac{q^m-1}{q+1}(1+ri)\ (\textrm{mod}\ \frac{q^{2m}-1}{q+1})\label{new last j}.
\end{eqnarray}
Let
\begin{eqnarray}\label{new i equivalent}
i=\frac{q^m-1}{q+1}w+\theta,
\end{eqnarray} where $w\in \Z, w\geq0$, and $0\leq\theta<\frac{q^m-1}{q+1}$. Then substituting Eq.(\ref{new i equivalent}) into Eq.(\ref{new last j}), we have
\begin{eqnarray}\label{new j equivalent}
\bar{i} \equiv q^m\theta+(1-w)\frac{q^m-1}{q+1} \ (\textrm{mod}\ t),
\end{eqnarray} where $t=\frac{q^{2m}-1}{q+1}$.
Since $0\leq i<\bar{i} \leq \delta^{max}-2$, there exists a positive integer $\overline{w}$, such that
$\frac{q^m-1}{q+1}\overline{w}\leq \delta^{max}-2 <\frac{q^m-1}{q+1}(\overline{w}+1)$. Thus, we have $0\leq w \leq \overline{w}\leq q-1$. We now consider three cases $w=0$, $w=1$, and $1<w \leq q-1$ $(q\geq 3)$, separately.\\
(1). For the case $w=0$, we have $i=\theta$, $\bar{i} \equiv q^m\theta+\frac{q^m-1}{q+1} \ (\textrm{mod}\ t)$, where $0\leq\theta<\frac{q^m-1}{q+1}$. It is easy to get that
\begin{eqnarray}\label{new case1}\left\{\begin{array}{ll}
0<\bar{i}=\frac{q^m-1}{q+1}\leq \delta^{max}-2<t, & \theta=0,\\
0<\delta^{max}-2<\bar{i}=q^m\theta+\frac{q^m-1}{q+1}<t,& \
0<\theta<\frac{q^m-1}{q+1}.
\end{array}\right.
\end{eqnarray}
This case means that for Eq.(\ref{m even equ7}), if $0<i=\theta<\frac{q^m-1}{q+1}$, then $\delta^{max}-2<\bar{i}<t$. And if $i=0$, then $0<\bar{i}=\frac{q^m-1}{q+1}\leq\delta^{max}-2$, which also implies $\bar{i}\equiv q-1\ (\textrm{mod}\ q^2)$.\\
(2). For the case $w=1$, we have $i=\frac{q^m-1}{q+1}+\theta$, $\bar{i} \equiv q^m\theta\ (\textrm{mod}\ t)$, where $0\leq\theta<\frac{q^m-1}{q+1}$.
It is easy to get that
\begin{eqnarray}\label{new case2}\left\{\begin{array}{ll}
\bar{i}=0, & \theta=0,\\
0<\delta^{max}-2<\bar{i}=q^m\theta<t,& \
0<\theta<\frac{q^m-1}{q+1}.
\end{array}\right.
\end{eqnarray}
This case implies that for Eq.(\ref{m even equ7}), if $i=\frac{q^m-1}{q+1}$, then $\bar{i}=0<i$, and if $i=\frac{q^m-1}{q+1}+\theta$,
$0\leq\theta<\frac{q^m-1}{q+1}$, then $\delta^{max}-2<\bar{i}<t$.\\
(3). For the case $1<w \leq q-1$ ($q\geq 3$), we have $i=\frac{q^m-1}{q+1}w+\theta$, where $0\leq\theta<\frac{q^m-1}{q+1}$. And
\begin{eqnarray}\label{case3j}\left\{\begin{array}{ll}
\bar{i}\equiv t-(w-1)\frac{q^m-1}{q+1}\ (\textrm{mod}\ t), & \theta=0,\\
\bar{i}\equiv q^m\theta-(w-1)\frac{q^m-1}{q+1}\ (\textrm{mod}\ t),& \
0<\theta<\frac{q^m-1}{q+1}.
\end{array}\right.
\end{eqnarray}
It is easy to get that
\begin{eqnarray}\label{case3}\left\{\begin{array}{ll}
\delta^{max}\!-\!2<t-(\overline{w}-1)\frac{q^m-1}{q+1}\leq\bar{i}=t-(w-1)\frac{q^m-1}{q+1}\leq t-\frac{q^m-1}{q+1}<t, & \theta=0,\\
\delta^{max}\!-\!2\!<\!2q^m\!-\!(\overline{w}\!-\!1)\frac{q^m\!-\!1}{q\!+\!1}\leq\! \bar{i}\!=\!q^m\theta\!-\!(w\!-\!1)\frac{q^m\!-\!1}{q\!+\!1}< \frac{q^m(q^m\!-\!1)}{q\!+\!1}\!-\!\frac{q^m\!-\!1}{q\!+\!1}\!<\!t,& 2\leq\theta<\frac{q^m-1}{q+1}.
\end{array}\right.
\end{eqnarray}
This case implies that for Eq.(\ref{m even equ7}), if $i=\frac{q^m-1}{q+1}w+\theta$, where $1<w \leq q-1$ $(q\geq 3)$,
$0\leq\theta<\frac{q^m-1}{q+1}$ and $\theta\neq1$, then $\delta^{max}-2<\bar{i}<t$.
For $\theta=1$, we have $i=\frac{q^m-1}{q+1}w+1$, and $\bar{i}=q^m-(w-1)\frac{q^m-1}{q+1}$, where $1<w\leq q-1 (q\geq 3)$.

By Eq.(\ref{m even equ6}) and the above discussions of three cases, we can infer that the eligible values of $i$, $\bar{i}$ satisfying $C_{1+ri}=C_{1+r\bar{i}}$
should hold
\begin{eqnarray}\label{eligible 1}
\bar{i}\equiv q-1\ (\textrm{mod}\ q^2),
\end{eqnarray} or
\begin{eqnarray}\label{eligible 2}
i=\frac{q^m-1}{q+1}w+1, \ \textrm{and}\ \bar{i}=q^m-(w-1)\frac{q^m-1}{q+1},
\end{eqnarray} where
$0\leq i < \bar{i}\leq \delta-2\leq \delta^{max}-2, 1<w \leq q-1 (q\geq 3)$.

Inversely, if $\bar{i}\equiv q-1\ (\textrm{mod}\ q^2)$, namely, $\bar{i}=q-1+kq^2$, where $k\in \Z, k\geq 0$, and $\bar{i}\leq \delta^{max}-2$ , then
$1+r\bar{i}\equiv q^2(1+rk)\ (\textrm{mod}\ q^{2m}-1)$. Thus, we let $i=k$, it is easy to get that $0\leq i < \bar{i}$, and $C_{1+ri}=C_{1+r\bar{i}}$. If
$\bar{i}=q^m-(w-1)\frac{q^m-1}{q+1}$, $i=\frac{q^m-1}{q+1}w+1$, where
$0\leq i < \bar{i}\leq \delta-2\leq \delta^{max}-2, 1<w \leq q-1 (q\geq 3)$, then $1+r\bar{i}\equiv q^m(1+ri)\ (\textrm{mod}\ q^{2m}-1)$, we also
have $C_{1+ri}=C_{1+r\bar{i}}$. Therefore, $C_{1+ri}=C_{1+r\bar{i}}$, $0\leq i < \bar{i}\leq \delta^{max}-2$, if and only if Eq.(\ref{eligible 1})
and Eq.(\ref{eligible 2}) hold.

For Eq.(\ref{eligible 1}), the number of the eligible $\bar{i}$, denoted by $N_{1}$, can be easily calculated by
$N_{1}=\lfloor\frac{\delta-2-(q-1)}{q^2}\rfloor+1=\lfloor\frac{\delta-(q+1)}{q^2}\rfloor+1$.
Let $N_{2}$ be the the number of the eligible $\bar{i}$ in Eq.(\ref{eligible 2}). In terms of Eq.(\ref{eligible 2}),
we have
\begin{eqnarray}\label{hold1}\left\{\begin{array}{ll}
\frac{q^m-1}{q+1}w+1<q^m-(w-1)\frac{q^m-1}{q+1},\\
q^m-(w-1)\frac{q^m-1}{q+1}\leq \delta-2,\\
2\leq w \leq q-1,\\
2\leq\delta\leq\delta^{max},
\end{array}\right.
\end{eqnarray}
namely,
\begin{eqnarray}\label{hold2}\left\{\begin{array}{ll}
q+2+\frac{(q+1)(3-\delta)}{q^m-1}\leq w\leq\lfloor\frac{q+1}{2}\rfloor,\\
2\leq\delta\leq\delta^{max}.
\end{array}\right.
\end{eqnarray}
It is not difficult to derive that by Eq.(\ref{hold2}), $w$ has integer solutions if and only if $\delta\geq3+\frac{(q^{m}-1)\lceil\frac{q+3}{2}\rceil}{q+1}=i^{*}\!+\!2+\frac{q^m-1}{q+1}$, and $q\geq 5$.
The number of the solutions of $w$ in Eq.(\ref{hold2}) can be calculated by
$\lfloor\lfloor\frac{q+1}{2}\rfloor-(q+2)-\frac{(q+1)(3-\delta)}{q^{m}-1}\rfloor+1$=
$\lfloor\frac{(q+1)(\delta-3)}{q^{m}-1}-\lceil\frac{q+3}{2}\rceil\rfloor$+1=$\lfloor\frac{(q+1)(\delta-3)}{q^{m}-1}-\lceil\frac{q+1}{2}\rceil\rfloor$.
Therefore, $N_{2}$ can be given by
\begin{eqnarray}\label{eqequihold1}N_{2}=\left\{\begin{array}{ll}
\lfloor\frac{(q+1)(\delta-3)}{q^{m}-1}-\lceil\frac{q+1}{2}\rceil\rfloor,&i^{*}\!+\!2+\frac{q^m-1}{q+1}\leq\delta\leq\delta^{max}, q\geq 5, \\
0, & otherwise.
\end{array}\right.
\end{eqnarray}
Combined with Lemma \ref{s cardinality} and Remark 1, we obtain the desired conclusions.\qed
By lemma \ref{exact cardinalities}, we can easily provide values of $\mathscr{N}(q,m,\delta)$ with specific $q$ and $m$ as follows.
\begin{corollary}\label{specific values}
Let the symbols be defined as above. Then we have\\
1. $\mathscr{N}(q\leq4,m=2,\delta)=2(\delta-N_{1}-1)$, where $2\leq\delta\leq\delta^{max}=\frac{q^3-q^2+q+3}{q+1}$.\\
2. $\mathscr{N}(q\leq4,m>2,\delta)=\left\{\begin{array}{ll}
(\delta-N_{1}-1)m,&\ 2\leq \delta<i^{*}+2,\\
(\delta-N_{1}-1-\frac{1}{2}[q=4])m,&\ i^{*}+2\leq \delta\leq \delta^{max}=\frac{q^{m+1}-q^2+q+3}{q+1}.\\
\end{array}\right.$\\
3. $\mathscr{N}(q\geq5\ odd,m,\delta)=\left\{\begin{array}{ll}
(\delta-N_{1}-1)m,&\ 2\leq \delta<i^{*}+2+\frac{q^{m}-1}{q+1},\\
(\delta-N_{1}-N_{2}-1)m,&\ i^{*}+2+\frac{q^{m}-1}{q+1}\leq \delta\leq \delta^{max}.\\
\end{array}\right.$\\
4. $\mathscr{N}(q\geq5\ even,m,\delta)=\left\{\begin{array}{ll}
(\delta-N_{1}-1)m,&\ 2\leq \delta<i^{*}+2,\\
(\delta-N_{1}-\frac{3}{2})m,&\ i^{*}+2\leq \delta<i^{*}+2+\frac{q^{m}-1}{q+1},\\
(\delta-N_{1}-N_{2}-\frac{3}{2})m,&\ i^{*}+2+\frac{q^{m}-1}{q+1}\leq \delta\leq \delta^{max}.\\
\end{array}\right.$
\end{corollary}
\emph{Proof}
Applying Lemma \ref{exact cardinalities}, the result follows. \qed
\subsection{Quantum codes construction and code comparisons }\label{Quantum Codes Construction}
In this subsection, we will construct quantum codes via the Hermitian construction. Firstly, we derive the parameters of narrow-sense constacyclic BCH codes as follows.
%\subsection{Construction of quantum codes from narrow-sense constacyclic BCH codes}
\begin{theorem}\label{classic Hermitian dual containing linear codes}
Let $n=\frac{q^{2m}-1}{q+1}$, where $q$ is a prime power, and $m\geq 2$ is even. Denote $\eta\in \mathbb{F}^*_{q^2}$, and ord$(\eta)=r=q+1$. Let $\mathcal{C}$ be a narrow-sense constacyclic BCH code with defining set $T=\bigcup_{i=0}^{\delta-2}C_{1+ri} $,
where $2 \leq \delta \leq \delta^{max}$.
Then $\mathcal{C}$ is an $[n,n-\mathscr{N}(q,m,\delta),\geq \delta]_{q^2}$ Hermitian dual-containing constacyclic BCH code,
where $\mathscr{N}(q,m,\delta)$ is given as in Lemma \ref{exact cardinalities}.
\end{theorem}
\emph{Proof}
Applying Theorem \ref{BCH bound} and Lemma \ref{exact cardinalities}, the result follows. \qed
\textbf{Remark 2}. We call a linear code with parameters $[n, k, d]_{q}$ optimal (almost optimal),
if the code (the code with parameters $[n, k, d+1]_{q}$) reaches some bound. Notice that some of constacyclic BCH codes constructed in
Theorem \ref{classic Hermitian dual containing linear codes} are optimal or almost optimal, they reach or almost reach the lower (upper)
bound maintained by Markus Grassl \cite{Grassl2020}. For example, by the Database \cite{Grassl2020}, the constacyclic BCH codes
$[5, 3, \geq3]_{2^{2}}$, $[20, 18, \geq2]_{3^{2}}$, $[20, 16, \geq4]_{3^{2}}$, $[20, 14, \geq5]_{3^{2}}$, $[85, 81, \geq3]_{2^{2}}$ are optimal
if the equalities of their minimum distance parameters are achieved. And the constacyclic BCH codes $[20, 12, \geq6]_{3^{2}}$, $[85, 77, \geq4]_{2^{2}}$, $[85, 73, \geq5]_{2^{2}}$, $[85, 69, \geq7]_{2^{2}}$, $[85, 65, \geq8]_{2^{2}}$
are almost optimal if the equalities of their minimum distance parameters are achieved, otherwise, they are optimal.

Now we construct quantum codes.
\begin{theorem}\label{quantum codes}
Let $n=\frac{q^{2m}-1}{q+1}$, where $q$ is a prime power, and $m\geq 2$ is even. Then there exists a quantum code with parameters $[[n, n-2\mathscr{N}(q,m,\delta), \geq \delta]]_{q}$,
where $\mathscr{N}(q,m,\delta)$ is given as in Lemma \ref{exact cardinalities}.
\end{theorem}
\emph{Proof}
Applying the Hermitian construction to the constacyclic code $\mathcal{C}$ in Theorem \ref{classic Hermitian dual containing linear codes}, we immediately obtain a $q$-ary quantum code with parameters $[[n, n-2\mathscr{N}(q,m,\delta),$ $\geq \delta]]_{q}$ as described in Theorem \ref{quantum codes}. \qed

Yuan et al. \cite{Yuan2017} constructed many quantum codes from a class of narrow-sense Hermitian dual containing constacyclic BCH codes
of length $n=\frac{q^{2m}-1}{q+1}$. Aly et al. in \cite{Aly2007} yielded quantum codes of general lengths by narrow-sense Hermitian dual containing BCH codes. Yves Edel provided code tables of some good quantum twisted codes for $q\leq 9$ and length up to 1000 \cite{edel2019}.
In \cite{ruihu2019}, Li et al. gave two classes of quantum codes from non-narrow-sense constacyclic codes, a class of which are with lengths
$n=\frac{(q-1)(q^{2}+1)}{b}$, where $b\mid q-1$.
Fixing the code lengths $n=\frac{q^{2m}-1}{q+1}$ ($m\geq 2$ even), we compare our quantum codes with known ones in \cite{Yuan2017, Aly2007, edel2019, ruihu2019}. Their results are provided as follows.
\begin{lemma}\label{yuan quantum codes}(\cite{Yuan2017}, Theorem 3)
Let $n=\frac{q^{2m}-1}{q+1}$, where $q$ is a prime power, and $m\geq 2$ is an even. Let $\alpha=1+(q+1)j_{\alpha}\in\{1+(q+1)i \mid 0\leq i \leq \delta_{Y}-2\}$, where $\delta_{Y}=\lfloor\frac{q}{q+1}(q^m-q+1)\rfloor+1$. Let $\lambda_{\alpha}=\frac{\alpha+q}{q+1}-\lfloor\frac{\alpha+q^3}{q^3+q^2}\rfloor$, then there exists a quantum code with parameters
\begin{eqnarray}\label{yuan2cyclotomic cosets cardinality}
\left\{\begin{array}{ll}
[[n,n-2\lambda_{\alpha}m,\geq j_{\alpha}+2]]_{q}, & \ q \ \textrm{is}\ \textrm{odd}, \ \textrm{or}\ q=2,\\
{[[n,n-2\lambda_{\alpha}m,\geq j_{\alpha}+2]]_{q}}, & \ q=2^{s}\ \textrm{with}\ s\geq 2,\ \alpha<(1\!+\!2^{s-1})(1\!+\!2^{sm}), \\
{[[n,n-2\lambda_{\alpha}m+m,\geq j_{\alpha}+2]]_{q}}, &\ \ q=4, m>2,\ \textrm{or}\ q=2^{s}\ \textrm{with}\ s\geq 2,\ \alpha \geq (1\!+\!2^{s-1})(1\!+\!2^{sm}).
\end{array}\right.
\end{eqnarray}
\end{lemma}
\begin{lemma}\label{aly quantum codes}(\cite{Aly2007}, Theorem 21)
Let $n=\frac{q^{2m}-1}{q+1}$, where $q$ is a prime power, $m=\textrm{ord}_{n}(q^2)\geq2$ even, and $2\leq\delta\leq\delta_{A}= \frac{q^m-1}{q+1}$,
then there exists a quantum code with parameters $[[n, n-2m\lceil(\delta-1)(1-q^{-2})\rceil,\geq \delta]]_{q}$.
\end{lemma}
\begin{lemma}\label{Li quantum codes odd}(\cite{ruihu2019}, Theorems 3 and 4)
Let $n=\frac{q^{4}-1}{q+1}$, where $q\geq5$ is an odd prime power. For $2\leq\delta\leq\delta_{L}=q^2-2q+3$, denote
$|T(\delta)|=2\lceil(\delta-2)(1-\frac{1}{2(q-1)})\rceil+1$.
Then there exists a quantum code with parameters $[[n, n-2|T(\delta)|,\geq \delta]]_{q}$.
\end{lemma}
\begin{lemma}\label{Li quantum codes even}(\cite{ruihu2019}, Theorem 5)
Let $n=\frac{q^{4}-1}{q+1}$, where $q\geq4$ is an even prime power. For $2\leq\delta\leq\delta_{L}=q^2-2q+3$, denote
\begin{eqnarray}
|T(\delta)|=\left\{\begin{array}{ll}2\lceil(\delta-2)(1-\frac{1}{2(q-1)})\rceil+1, & 2\leq\delta\leq q^2-3q+4,\\
2(\delta-2-\frac{q-2}{2})+1, & q^2-3q+5\leq\delta\leq q^2-2q+3,\end{array}\right.
\end{eqnarray}
Then there exists a quantum code with parameters $[[n, n-2|T(\delta)|,\geq \delta]]_{q}$.
\end{lemma}

By comparison, it shows that for $n=\frac{q^{2m}-1}{q+1}$ ($m$ is even), quantum codes constructed by Theorem \ref{quantum codes} have the same parameters with those in \cite{Yuan2017} with $q=2, 3, 4$.
Whereas, with the same lengths, if $q\geq5$, for all design distance $\delta$ in the range $i^{*}+2+\frac{q^m-1}{q+1}\leq\delta\leq\delta^{max}$, our quantum codes can always yield strict dimension or design distance gains than the ones presented in \cite{Yuan2017}. Tables 1, 2 provide some code comparisons between them. From these tables, we can see that our quantum codes have better performance. Moreover, for the maximum design distance $\delta^{max}$ in this paper, $\delta_{A}$ in \cite{Aly2007},
$\delta_{Y}$ in \cite{Yuan2017}, and $\delta_{L}$ in \cite{ruihu2019}, it is easy to infer that $\delta^{max}>(q-1)\delta_{A}$,
$\delta^{max}=\delta_{Y}\ \textrm{or}\  \delta_{Y}+1$, and $\delta^{max}=\delta_{L}\ \textrm{or}\ \delta_{L}+1$. It means that  for length $n=\frac{q^{2m}-1}{q+1}$ ($m$ is even),
the Hermitian dual containing constacyclic BCH codes studied in this paper have relatively large
design distance. We observe that quantum codes obtained by Theorem \ref{quantum codes} can yield better parameters than those known constructions of quantum BCH codes or good quantum twisted codes \cite{ edel2019, Aly2007}.
We also notice that non-narrow sense constacyclic BCH codes or BCH codes investigated in \cite{ruihu2019} can construct abundant quantum codes with good parameters. Nonetheless, with the same length, parameters of quantum codes constructed by non-narrow sense codes are not always superior to
the narrow-sense ones, some of quantum codes constructed from narrow-sense constacyclic BCH codes in this paper have better performance than the ones in \cite{ruihu2019}.
Since Yves Edel's table \cite{edel2019} includes good quantum twisted codes with lengths up to 1000, we just compare our quantum codes in Theorem \ref{quantum codes} with those obtained in Refs.\cite{ edel2019, ruihu2019, Aly2007} in the case of $n<1000$. By table 3, it is easy to see that for the same length, our quantum codes have
better code rate or larger design distance than the ones available in \cite{edel2019, ruihu2019, Aly2007}, where the symbol $"-"$ means that there is no quantum code with given length or design distance.

\textbf{Remark 3}. After we finish this paper, we find that Wang et al. in \cite{liqi2019} deal with a family of narrow-sense constacyclic BCH codes of length $\frac{q^{2m}-1}{\rho}$ ( $\rho$ is a positive divisor of ($q+1$), and $m\geq 3$), which extend the length $n=\frac{q^{2m}-1}{q+1}$ to more general case. Compared with the codes of length $n=\frac{q^{2m}-1}{q+1}$ in \cite{liqi2019}, our results have the following merits.
(1). For even $m$, we enlarge the range of $m$. Wang et al. in  \cite{liqi2019} discuss the cases $m\geq 4$,
we include the case $m=2$. So we can provide many short lengths Hermitian dual containing constacyclic BCH codes and related quantum codes, which will be interesting from a practical point of view. (2). We simplify the dimension expressions of constacyclic BCH codes or related quantum codes. In \cite{liqi2019}, for given design distance $\delta$, one should first get the $q$-adic expansion of $1+(q+1)(\delta-2)$
by $\Sigma_{i=0}^{m}\delta_{i}q^{i}$, and then discuss the values of $\delta_{i}$ ($0\leq i \leq m)$, thereby $\delta$ is the function of $q, m, \delta_{i}$ ($0\leq i \leq m$),  and the dimension of the Hermitian dual-containing constacyclic BCH code or related quantum code is the function of $q, m, \delta, \delta_{0}$, and $\delta_{m}$ (see Lemma 5 and Theorem 4 in \cite{liqi2019}). While
by our Theorems \ref{classic Hermitian dual containing linear codes} and
\ref{quantum codes}, the dimension of the Hermitian dual-containing constacyclic BCH codes is a function of their design
parameters, namely, $n, q$, and the given design distance $\delta$, as well as the related quantum codes. Quantum codes of length $n=\frac{q^{2m}-1}{q+1}$ also have been studied in \cite{Hao2018} using non-narrow-sense BCH codes, and many quantum codes in their paper have parameters better than our constructions from
narrow-sense constacyclic BCH codes. However, we relax the values of even $m$ from $m\geq4$ to $m\geq2$, and remove the condition $q\equiv1 (\textrm{mod}\ m)$. Thus, the obtained constacyclic BCH codes or resultant quantum codes in this paper are beneficial complements for the know ones with length $n=\frac{q^{2m}-1}{q+1}$.

\begin{table}[!htbp]\label{table1}
\centering
\caption{Quantum codes(QC) of length $n=\frac{q^{2m}-1}{q+1} $ with $m=2$, $q=5,7,8,9$}
\footnotesize
%\small
\begin{tabular}{|p{1.5cm}|p{2.16cm}|p{2.16cm}|p{2.16cm}|p{2.16cm}|}
\hline
$m$, $q$, $\delta$ &QC in Theorem \ref{quantum codes}& QC in \cite{Yuan2017}  &  QC in Theorem \ref{quantum codes} & QC in \cite{Yuan2017}\\
\hline
$m=2, q=5$           & $\cdots$   &    &       & \\
$\delta=19$    &$[[104,40,\geq 19]]_{5}$  & $[[104,40,\geq 18]]_{5}$ &    &  \\

  $m=2, q=7$         & $\cdots$   &    &       & \\
 $33\leq\delta\leq39$    &$[[300,180,\geq 33]]_{7}$  & $[[300,176,\geq 33]]_{7}$ & $[[300,176,\geq 34]]_{7}$ & $[[300,172,\geq 34]]_{7}$ \\

                     &$[[300,172,\geq 35]]_{7}$  & $[[300,168,\geq 35]]_{7}$ & $[[300,168,\geq 36]]_{7}$ & $[[300,164,\geq 36]]_{7}$\\

                      &$[[300,164,\geq 37]]_{7}$  & $[[300,160,\geq 37]]_{7}$ & $[[300,160,\geq 39]]_{7}$ & $[[300,156,\geq 38]]_{7}$\\

$m=2, q=8$  & $\cdots$   &    &       & \\
$45\leq\delta\leq52$     &$[[455,289,\geq 45]]_{8}$  & $[[455,285,\geq 45]]_{8}$ & $[[455,285,\geq 46]]_{8}$  & $[[455,281,\geq 46]]_{8}$ \\

                      &$[[455,281,\geq 47]]_{8}$  & $[[455,277,\geq 47]]_{8}$  & $[[455,277,\geq 48]]_{8}$  & $[[455,273,\geq 48]]_{8}$\\

                      &$[[455,273,\geq 49]]_{8}$  & $[[455,269,\geq 49]]_{8}$ & $[[455,269,\geq 50]]_{8}$  & $[[455,265,\geq 50]]_{8}$\\

                      &$[[455,265,\geq 52]]_{8}$  & $[[455,261,\geq 51]]_{8}$ &    & \\

$m=2, q=9$& $\cdots$   &    &       & \\
  $51\leq\delta\leq67$  &$[[656,464,\geq 51]]_{9}$  & $[[656,460,\geq 51]]_{9}$   & $[[656,460,\geq 52]]_{9}$   & $[[656,456,\geq 52]]_{9}$ \\

   &$[[656,456,\geq 53]]_{9}$   & $[[656,452,\geq 53]]_{9}$ & $[[656,452,\geq 54]]_{9}$   & $[[656,448,\geq 54]]_{9}$\\

                      &$[[656,448,\geq 55]]_{9}$   & $[[656,444,\geq 55]]_{9}$ & $[[656,444,\geq 56]]_{9}$   & $[[656,440,\geq 56]]_{9}$\\

                      &$[[656,440,\geq 57]]_{9}$   & $[[656,436,\geq 57]]_{9}$ & $[[656,436,\geq 58]]_{9}$   & $[[656,432,\geq 58]]_{9}$\\

                      &$[[656,436,\geq 59]]_{9}$   & $[[656,428,\geq 59]]_{9}$ & $[[656,432,\geq 60]]_{9}$   & $[[656,424,\geq 60]]_{9}$\\

                      &$[[656,428,\geq 61]]_{9}$   & $[[656,420,\geq 61]]_{9}$ & $[[656,424,\geq 62]]_{9}$   & $[[656,416,\geq 62]]_{9}$\\

                      &$[[656,420,\geq 63]]_{9}$   & $[[656,412,\geq 63]]_{9}$ & $[[656,416,\geq 64]]_{9}$   & $[[656,408,\geq 64]]_{9}$\\

                      &$[[656,412,\geq 65]]_{9}$   & $[[656,404,\geq 65]]_{9}$ & $[[656,408,\geq 67]]_{9}$   & $[[656,400,\geq 66]]_{9}$\\
$\cdots$  & $\cdots$ & $\cdots$   & $\cdots$  & $\cdots$ \\
\hline
\end{tabular}
%\end{center}
\end{table}
\begin{table}[!htbp]\label{table1}
\centering
\caption{Quantum codes(QC) of length $n=\frac{q^{2m}-1}{q+1} $ with $m=4$, $q=5,7$}
\footnotesize
%\small
\begin{tabular}{|p{3.2cm}|p{3.9cm}|p{3.9cm}|}
\hline
$m$, $q$, $\delta$ &QC in Theorem \ref{quantum codes}& QC in \cite{Yuan2017}  \\
\hline
$m=4, q=5$            &$[[65104,61904,\geq 419]]_{5}$  & $[[65104,61896,\geq 419]]_{5}$   \\

$419\leq\delta\leq518 $ & $[[65104,61896,\geq 420]]_{5}$  & $[[65104,61888,\geq 420]]_{5}$ \\

$ $ & $\cdots $  & $\cdots $ \\

           &$[[65104,61144,\geq 518]]_{5}$   & $[[65104,61136,\geq 518]]_{5}$   \\

$m=4, q=7$            &$[[720600,708840,\geq 1503]]_{7}$  & $[[720600,708832,\geq 1503]]_{7}$   \\

$1503\leq\delta\leq2096 $ & $[[720600,708832,\geq 1504]]_{7}$  & $[[720600,708824,\geq 1504]]_{7}$  \\

$ $ & $\cdots $  & $\cdots $ \\

          & $[[720600,704200,\geq 2096]]_{7}$   & $[[720600,704184,\geq 2096]]_{7}$  \\
$\cdots $ & $\cdots $  & $\cdots $ \\
\hline
\end{tabular}
%\end{center}
\end{table}

\begin{table}[!htbp]\label{table1}
\centering
\caption{Quantum codes(QC) of length $n=\frac{q^{2m}-1}{q+1}\ (n<1000)$}
%\footnotesize
\small
\begin{tabular}{|p{1.2cm}|p{2.3cm}|p{2.3cm}|p{2.2cm}|p{1.9cm}|}
\hline
m,\  q & QC in Theorem \ref{quantum codes}& QC in \cite{ruihu2019} & QC in \cite{edel2019}  & QC in \cite{Aly2007}\\
\hline
m=2, q=2  & $[[5,1,\geq 3]]_{2}$ & - &  - & \\

m=2, q=3  & $[[20,12,\geq 4]]_{3}$ & $[[20,10,\geq 4]]_{3}$ &  $[[20,12,3]]_{3}$ & \\

          & $[[20,8,\geq 5]]_{3}$ & - & - & \\

  %       & $[[20,6,\geq 6]]_{3}$ & $[[20,4,\geq 6]]_{3}$ &  & \\

m=2, q=4  & $[[51,39,\geq 5]]_{4}$  & $[[51,37,\geq 5]]_{4}$  & $[[51,39,5]]_{4}$ &\\

          & $[[51,35,\geq 6]]_{4}$  & $[[51,33,\geq 6]]_{4}$  & $[[51,35,6]]_{4}$ & \\

          & $[[51,31,\geq 7]]_{4}$  &  -  & $[[51,31,7]]_{4}$ & \\

m=4,  q=2  & $[[85,77,\geq 3]]_{2}$ & &$[[85,77,3]]_{2}$ & $[[85,77,\geq 2]]_{2}$\\

           & $[[85,69,\geq 4]]_{2}$ & &$[[85,69,4]]_{2}$ & $[[85,69,\geq 3]]_{2}$\\

           & $[[85,61,\geq 5]]_{2}$ & &$[[85,61,5]]_{2}$ & $[[85,61,\geq 5]]_{2}$\\

           & $[[85,53,\geq 7]]_{2}$ & &$[[85,53,7]]_{2}$ & - \\

          % & $[[85,45,\geq 8]]_{2}$ & &$[[85,49,8]]_{2}$ & - \\

          % & $[[85,37,\geq 9]]_{2}$ & &$[[85,41,8]]_{2}$ & - \\

           & $[[85,29,\geq 11]]_{2}$ & & - & - \\

m=2,  q=5  & $[[104,88,\geq 6]]_{5}$ &$[[104,86,\geq 6]]_{5}$ &$[[104,86,6]]_{5}$ & \\

           & $[[104,84,\geq 7]]_{5}$ &$[[104,82,\geq 7]]_{5}$ & $[[104,82,7]]_{5}$ & \\

           & $[[104,80,\geq 8]]_{5}$ &$[[104,78,\geq 8]]_{5}$ & $[[104,78,8]]_{5}$ & \\

           & $[[104,76,\geq 9]]_{5}$ &- &  -                  & \\

           & $[[104,40,\geq 19]]_{5}$ &- &  -                 & \\

m=2, q=7  & $[[300,276,\geq 8]]_{7}$ &$[[300,274,\geq 8]]_{7}$ &  $[[300,274,8]]_{7}$ & \\

          & $[[300,272,\geq 9]]_{7}$ &$[[300,270,\geq 9]]_{7}$ &  $[[300,270,9]]_{7}$ & \\

          & $[[300,268,\geq 10]]_{7}$ &$[[300,266,\geq 10]]_{7}$ &  $[[300,266,10]]_{7}$ & \\

          & $[[300,264,\geq 11]]_{7}$ &$[[300,262,\geq 11]]_{7}$ &  $[[300,262,11]]_{7}$ & \\

          & $[[300,260,\geq 12]]_{7}$ &$[[300,258,\geq 12]]_{7}$ &  $[[300,258,12]]_{7}$ & \\

          & $[[300,256,\geq 13]]_{7}$ &- & -  & \\

          & $[[300,160,\geq 39]]_{7}$ &- & -  & \\

m=2, q=8  & $[[455,427,\geq 9]]_{8}$  &  $[[455,425,\geq 9]]_{8}$   &  & \\

          & $[[455,423,\geq 10]]_{8}$ &  $[[455,421,\geq 10]]_{8}$   &  & \\

          & $[[455,419,\geq 11]]_{8}$ &  $[[455,417,\geq 11]]_{8}$   &  & \\

          & $[[455,415,\geq 12]]_{8}$ &  $[[455,413,\geq 12]]_{8}$   & & \\

          & $[[455,411,\geq 13]]_{8}$ &  $[[455,409,\geq 13]]_{8}$   &  & \\

          & $[[455,407,\geq 14]]_{8}$ &  $[[455,405,\geq 14]]_{8}$   &  & \\

          & $[[455,403,\geq 15]]_{8}$ &  -                            &  & \\

          & $[[455,265,\geq 52]]_{8}$ &  -                            &  & \\

m=2, q=9 & $[[656,624,\geq 10]]_{9}$ &$[[656,622,\geq 10]]_{9}$ & $[[656,622, 10]]_{9}$  & \\

         & $[[656,620,\geq 11]]_{9}$ &$[[656,618,\geq 11]]_{9}$& $[[656,618, 11]]_{9}$ & \\

         & $[[656,616,\geq 12]]_{9}$ &$[[656,614,\geq 12]]_{9}$ & $[[656,614, 12]]_{9}$ & \\

         & $[[656,612,\geq 13]]_{9}$ &$[[656,610,\geq 13]]_{9}$ & $[[656,610, 13]]_{9}$ & \\

         & $[[656,608,\geq 14]]_{9}$ &$[[656,606,\geq 14]]_{9}$  & $[[656,606, 14]]_{9}$ & \\

         & $[[656,604,\geq 15]]_{9}$ &$[[656,602,\geq 15]]_{9}$ & $[[656,602, 15]]_{9}$ & \\

         & $[[656,600,\geq 16]]_{9}$ &$[[656,598,\geq 16]]_{9}$ & $[[656,598, 16]]_{9}$ & \\

         & $[[656,596,\geq 17]]_{9}$ &- &-   & \\

         & $[[656,408,\geq 67]]_{9}$ &- &-   & \\
\hline
\end{tabular}
%\end{center}
\end{table}
\section{Conclusion}\label{Summary}
In this paper, the maximum design distance and properties of the corresponding cyclotomic cosets for a family of narrow-sense Hermitian dual-containing constacyclic BCH codes $\mathcal{C}$ with length $\frac{q^{2m}-1}{q+1}$ ($m\geq 2$ is an even) were deeply explored. Furthermore, the exact dimension of $\mathcal{C}$ with design distance $\delta$ in the rang $2\leq \delta\leq \delta^{max}$ was completely determined. Consequently, applying Hermitian construction to these underlying constacyclic BCH codes, many quantum codes with desirable parameters were constructed.

\begin{acknowledgements}
This work is supported by the National Natural Science Foundation of China under Grant No.61902429, No.11775306, the Shandong Provincial Natural Science Foundation of China under Grants No.ZR2019MF070, the Key Laboratory of Applied Mathematics of Fujian Province University (Putian University) under Grants No.SX201806, the Open Research Fund from Shandong provincial Key Laboratory of Computer Network under Grant No.SDKLCN-2018-02,
Fundamental Research Funds for the Central Universities No.17CX02030A.
\end{acknowledgements}

%\begin{acknowledgements}
%\end{acknowledgements}




\end{document}